\documentclass[lettersize,journal]{IEEEtran}
\usepackage{amsmath,amsfonts}
\usepackage{algorithmic}
\usepackage{algorithm}
\usepackage{array}
\usepackage[caption=false,font=normalsize,labelfont=sf,textfont=sf]{subfig}
\usepackage{textcomp}
\usepackage{stfloats}
\usepackage{url}
\usepackage{graphicx}
\usepackage{cite}
\usepackage[hidelinks]{hyperref}
\usepackage{booktabs}
\usepackage{multirow}
\begin{document}

\title{Saturation-Aware Predictive Quantization for Low-Power ECG Acquisition:\\A Benchmark of Taylor, Adaptive-Order, Kalman, and LSTM Predictors}

\author{
    % 1. 作者名字，并用 \textsuperscript{*} 手动添加右上角星号
    Xiyuan~Feng\textsuperscript{1},
    Yuxiang~Zhao\textsuperscript{1},
    Jie~Xiong\textsuperscript{1},
    Dian~Lin\textsuperscript{1},
    Yunlei~Zhong\textsuperscript{2},
    Wei~Liu\textsuperscript{1},
    Zhongheng~Ji\textsuperscript{1},
    Ruiyu~Tian\textsuperscript{1},
    Chenhao~Zhuo\textsuperscript{1,*},
    and~Yue~Yin\textsuperscript{1,*}
    
    % 2. 强制换行，并在下方显示地址（使用 \small 缩小字号，使其美观）
    \and [0.5em] \small \textsuperscript{1}School of Integrated Circuits (School of Microelectronics), Northwestern Polytechnical University, 
    No. 1 Dongxiang Road, Chang’an District, Xi’an 710129, P. R. China.
    
    \and [0.5em] \small \textsuperscript{2}Key Laboratory of Multifunctional Nanomaterials and Smart Systems Division of Advanced Materials,Suzhou
Institute of Nano-Tech and Nano-Bionics, Chinese Academy of Sciences, 
     Suzhou, 215128, , P. R. China
    % 3. 再次强制换行，显示通讯作者及邮箱
    
    \and [0.3em] \small *Corresponding authors: Yue Yin (yinyue@nwpu.edu.cn) and Chenhao Zhuo (zhuochenhao@outlook.com)
    
    % 4. 传统的基金和时间信息依然可以通过 \thanks 留在左下角脚注（如果不需要脚注，可以把下面这几行删掉）
    \thanks{This paper was produced by the IEEE Publication Technology Group. They are in Piscataway, NJ.}
    \thanks{Manuscript received April 19, 2021; revised August 16, 2021.}
    \thanks{This work was also supported in part by the Fundamental Research Funds for the Central Universities of Ministry of Education of China (No. D5000240188) and a fellowship from the China Postdoctoral Science Foundation (No. 2025M784413).}
}
\markboth{IEEE Transactions on Biomedical Circuits and Systems}%
{Author \MakeLowercase{\textit{et al.}}: Predictive Quantization for Low-Power ECG Acquisition}

\maketitle

\begin{abstract}
Wearable electrocardiogram (ECG) monitors require energy-efficient analog-to-digital converters (ADCs), yet conventional successive-approximation-register (SAR) ADCs repeatedly resolve slowly varying most significant bits. Predictive quantization (PQ) instead estimates the next sample and quantizes only the residual, thereby reducing the required conversion depth. Its principal failure mode is residual saturation, which occurs when prediction error exceeds the residual ADC range and is irreversibly clipped. We compared four one-step-ahead predictors under a common 10-bit, saturation-aware PQ model with residual widths from 2 to 8 bits. The benchmark included first-order Taylor extrapolation, an adaptive-order predictor, a constant-velocity Kalman filter, and a two-layer long short-term memory (LSTM) network. We used an open-loop protocol in which all predictors received past original samples. This protocol isolates intrinsic prediction performance from recursive reconstruction-error propagation. Saturation rate (SR) was the primary metric, complemented by overflow energy ratio (OER), which weights each event by its squared overflow depth. On a 5,317-sample excerpt from MIT-BIH Arrhythmia Database Record 101, the Kalman predictor performed best at $B_r=6$. It achieved 30.88\,dB SNR, 2.16\% SR, and 12.93\% OER, compared with 28.39\,dB, 2.69\%, and 25.29\% for Taylor extrapolation. The adaptive-order predictor achieved 29.61\,dB SNR and 2.44\% SR using three registers, two comparators, and no multiplier. The LSTM reached 29.28\,dB SNR and did not outperform the model-based predictors on this limited-data benchmark. Under the evaluated excerpt and open-loop protocol, $B_r=6$ provided a favorable balance between reconstruction fidelity and conversion depth. Closed-loop, multi-subject, and hardware validation are required before system-level energy or deployment claims can be made.
\end{abstract}


% ======================================================================
\section{Introduction}
% ======================================================================

\IEEEPARstart{W}{earable} ECG monitors support continuous rhythm assessment outside clinical settings, but their operating lifetime is constrained by a limited battery budget. The analog front-end, including the ADC, therefore requires aggressive energy optimization during continuous acquisition.

\subsection{Traditional ADC Power Bottleneck}

SAR ADCs are widely used for wearable biosignal acquisition because they are efficient at moderate resolution and sampling rates. A conventional SAR converter resolves each code through sequential comparisons, beginning with the MSB and switching a binary-weighted capacitor array at each step. The largest capacitors are associated with the most energy-intensive decisions \cite{ref4,ref5}. This conversion sequence is poorly matched to oversampled ECG signals, whose adjacent samples are strongly correlated and whose MSBs change infrequently. Nevertheless, a conventional SAR ADC resolves every bit during every conversion.

\subsection{Predictive Quantization: Principle and Promise}

Predictive quantization exploits this temporal redundancy by quantizing the prediction residual $r[n]=x[n]-\hat{x}[n]$ rather than the full input $x[n]$. Here, $\hat{x}[n]$ is a one-step-ahead estimate derived from previous samples. An accurate predictor concentrates the residual within a smaller range, allowing the converter to use fewer bit trials per sample. We denote the residual width by $B_r$ and the nominal ADC resolution by $N_{\text{total}}$. Their ratio, $N_{\text{total}}/B_r$, is a conversion-depth reduction factor rather than a direct measurement of energy savings. For a 10-bit ADC with $B_r=6$, the predictor supplies four MSBs and the SAR ADC resolves the six-bit residual.

\begin{figure}[!t]
\centering
\includegraphics[width=3.3in]{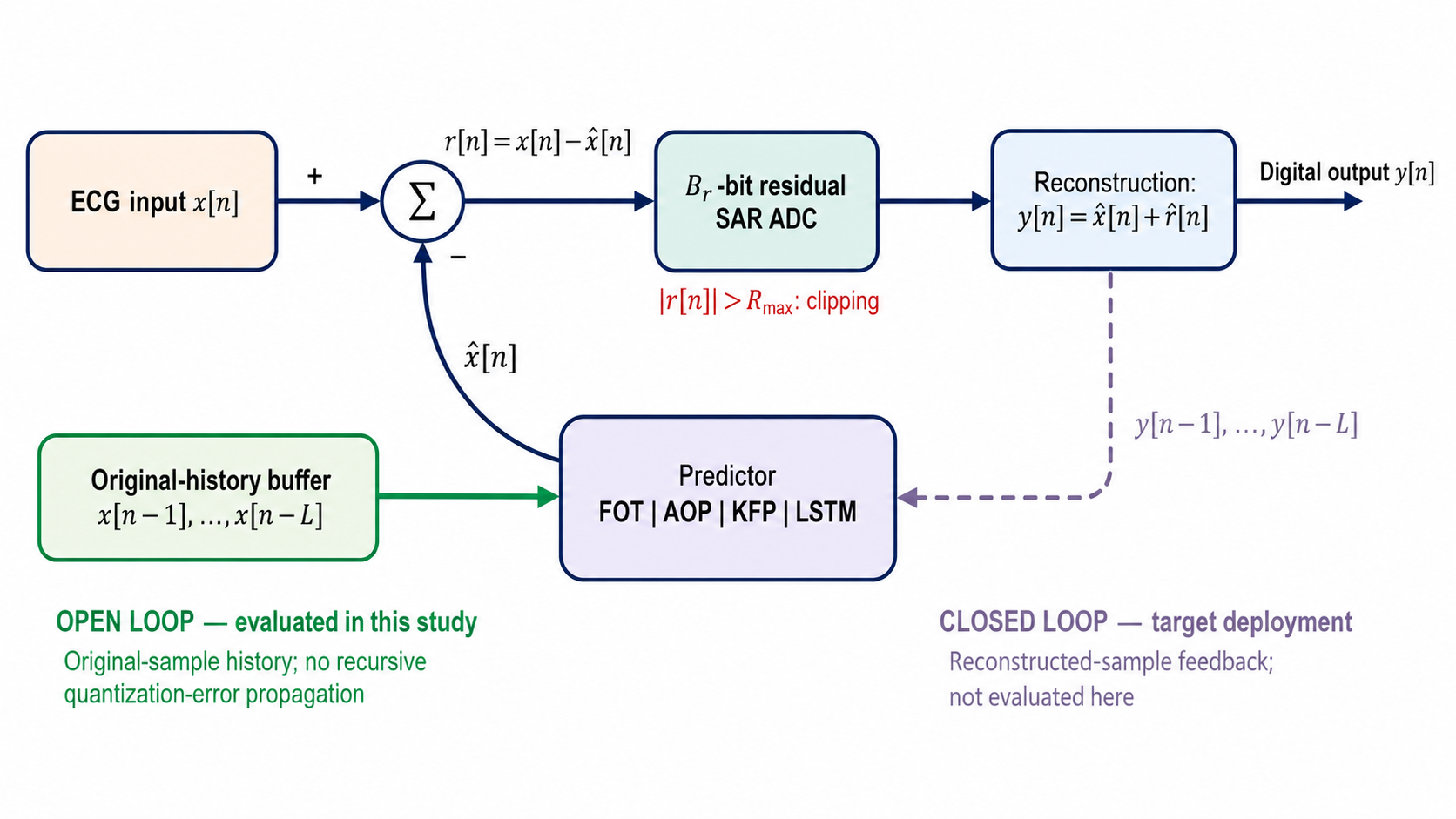}
\caption{Predictive quantization ADC architecture with open-loop evaluation and closed-loop deployment paths. 
The predictor generates $\hat{x}[n]$ and provides the upper $(10-B_r)$ bits, while the $B_r$-bit residual SAR ADC quantizes only the residual $r[n]=x[n]-\hat{x}[n]$. 
Residuals exceeding $\pm R_{\max}$ are clipped before reconstruction. 
The green solid path denotes the open-loop benchmark evaluated in this work using original-sample history, avoiding recursive quantization-error propagation. 
The purple dashed path indicates the target closed-loop architecture using reconstructed-sample feedback, which is not evaluated in the present study.}
\label{fig:system}
\end{figure}

Figure~\ref{fig:system} shows the target closed-loop architecture. At sample $n$, the predictor generates $\hat{x}[n]$, and the ADC quantizes the residual $r[n]$ to obtain $\hat{r}[n]$. The reconstructed sample, $y[n]=\hat{x}[n]+\hat{r}[n]$, is then returned to the predictor in a deployed system. The present benchmark instead uses past original samples to isolate one-step prediction performance. Residual saturation occurs when $|r[n]|$ exceeds the ADC input range, forcing hard clipping and irreversible reconstruction error.

\subsection{Related Work}

The concept of reducing ADC conversion energy by exploiting temporal signal redundancy spans a rich design space, from predictive algorithms that skip conversion cycles to learned models that anticipate waveform morphology. We survey this landscape along four axes: predictive ADC architectures, data-dependent conversion algorithms, MSB prediction schemes, and ECG prediction models.

\textbf{Predictive ADC architectures.} The idea of equipping an ADC with a predictor that supplies an initial estimate, so that only the residual requires conversion, was formalized by Wood and Sun~\cite{ref14} as the ``Predicting ADC.'' They demonstrated that a first-order difference predictor can reduce the number of required comparisons per sample, yielding both speed and power benefits. Mitrovic et al.~\cite{ref23} extended this concept to a general predictive successive-approximation framework, analyzing the tradeoff between predictor accuracy and the probability of out-of-range errors that necessitate fallback conversion cycles. Scanlan et al.~\cite{ref20} provided a system-level analysis of feedback predictive-encoder-based ADCs, modeling the predictor, quantizer, and reconstruction loop as a unified control system.

The predictive ADC concept is particularly compelling for biomedical signals, whose strong temporal correlation makes them highly predictable over short horizons. Van Rethy et al.~\cite{ref13} were among the first to demonstrate a predictive-sensing SAR ADC specifically targeting ECG signals: their architecture predicts the next sample and confines the SAR conversion to a subrange, achieving significant energy savings compared to a full-range conversion. Namavar et al.~\cite{ref15} reported a third-order linear predictive SAR ADC for neural recording implants, operating at 330\,nW with a 10-bit resolution in 180\,nm CMOS; the predictor uses three prior digital output codes to generate the estimate, reducing the average bit-cycles per conversion. Zhang et al.~\cite{ref21} proposed a dynamic tracking algorithm for bio-related SAR ADCs that maintains two adjacent tracking windows around the prediction interval, enabling the ADC to automatically adjust the subrange and update the prediction center as the input signal drifts. Feng et al.~\cite{ref1} combined predictive quantization with compressive sensing for ECG acquisition, achieving a 30\% reduction in average conversion depth. Tang et al.~\cite{ref2} introduced dynamic predictive sampling (DPS), where a linear extrapolator generates a tracking window and a full SAR conversion is triggered only when the input falls outside it, achieving $6.17\times$ data compression and 31\% power saving. Renteria-Pinon et al.~\cite{ref3} extended this to a fully digital second-order level-crossing ADC, selecting only fiducial turning points for quantization and achieving $8.33\times$ compression with on-chip ECG delineation.

These chip-scale demonstrations share a common premise---exploit signal predictability to skip conversion cycles---but employ predictors of limited complexity: linear extrapolation, polynomial fitting, or pseudo-random sequence correlation. None has evaluated nonlinear, learned, or probabilistically motivated predictors under a unified saturation model.

\textbf{Data-dependent and LSB-first algorithms.} An orthogonal approach to reducing ADC energy exploits the observation that when the input signal is slowly varying, the MSBs do not change from sample to sample and therefore need not be re-determined. Yaul and Chandrakasan~\cite{ref16} pioneered the LSB-first successive approximation algorithm, in which conversion begins from the LSB side and proceeds toward the MSB side only as needed; the number of bit-cycles scales logarithmically with signal activity rather than being fixed at $N$. Chen et al.~\cite{ref17} combined predictive estimation with LSB-first conversion, using the prediction to seed the initial LSB window and thereby accelerating sub-radix-2 SAR conversion. Lu et al.~\cite{ref20b} proposed a self-adaptive window algorithm that configures an $N$-bit LSB window based on signal characteristics and extends it toward the MSB side only when the signal leaves the current window. Inanlou et al.~\cite{ref24} introduced arithmetic tracking, where the ADC determines the required step size from the signal's recent activity and bypasses conversion cycles during low-activity periods. Zhang et al.~\cite{ref8b} applied a quantization-error-matched reverse-LSB-first algorithm to noise-shaping SAR ADCs, saving 87.3\% of switching power under oversampling.

\textbf{MSB prediction and bit-guess schemes.} Rather than restructuring the conversion sequence, several designs retain the conventional MSB-first order but use auxiliary circuits to predict the initial MSB values. Lin and Hsieh~\cite{ref18} proposed a first 2-bit guess (F2G) scheme that presets the two MSB capacitors based on the previous sample's output code, reducing DAC switching energy by 90\% in theory. Lin et al.~\cite{ref6b} extended this to an input range prediction DAC switching technique that narrows the SAR trial range to prevent unnecessary capacitor switching. Canal et al.~\cite{ref19} employed a time-to-digital converter (TDC) to predict the three MSB DAC capacitor switching values in a single SAR cycle, combined with correlated-reversed switching for linearity improvement. Zhang et al.~\cite{ref3b} proposed a fully-predictive ADC with a code-recombination algorithm that reuses the previous sample's code to skip the most significant bit-trials entirely. These hardware-level MSB prediction schemes are complementary to the algorithmic predictor evaluation in this work: they demonstrate the circuit feasibility of skipping MSB cycles, while our study addresses the question of \textit{which} predictor to use and \textit{how accurately} it must predict to keep residual saturation within acceptable bounds.

\textbf{Predictive noise-shaping.} Shakya et al.~\cite{ref22} recently combined prediction with first-order noise shaping in a SAR ADC, using the prediction to restrict the SAR conversion range while the noise-shaping loop suppresses quantization error. This work suggests a further dimension---the interaction between prediction accuracy and noise-shaping order---that remains unexplored.

\textbf{ECG prediction algorithms.} On the algorithm side, one-step-ahead ECG prediction has been studied independently of ADC design. Zacarias et al.~\cite{ref6} developed a two-layer LSTM model evaluated on all 48 records of the MIT-BIH Arrhythmia Database, reporting a mean absolute error of $0.0522 \pm 0.0098$. Wang et al.~\cite{ref7} proposed a Nested LSTM architecture achieving a 24.6\% RMSE reduction over standard LSTM on the PhysioNet SCD Holter Database. These studies evaluate prediction accuracy in isolation---via root-mean-square error (RMSE) or mean absolute error (MAE)---without modeling the hard saturation boundary that defines predictive ADC behavior.

\textbf{SAR ADC low-power design space.} At the architecture level, Arafa et al.~\cite{ref4} provided a comprehensive survey of SAR ADC designs for biomedical applications, identifying prediction-based conversion as an emerging direction. Tong and Ghovanloo~\cite{ref5} benchmarked capacitor switching schemes, demonstrating over 97\% switching energy reduction with advanced schemes---though these savings address DAC switching rather than the bit-trial elimination that PQ targets.

\textbf{Gap.} Prior studies have not compared predictors across this complexity range under a common PQ simulation with an explicit residual-saturation boundary. Moreover, SR alone does not distinguish shallow threshold crossings from deep overflow events. This study addresses both limitations through a matched predictor benchmark and overflow-sensitive metrics.

\subsection{Contributions}

This paper makes three contributions:

\begin{enumerate}
\item \textbf{A systematic four-predictor benchmark under predictive quantization.} We evaluate first-order Taylor extrapolation (FOT), an adaptive-order predictor (AOP) that switches between zeroth-, first-, and second-order modes based on local signal statistics, a constant-velocity Kalman filter predictor (KFP), and a two-layer stacked LSTM network on MIT-BIH Arrhythmia Database Record 101 (360\,Hz), with residual bit-widths of 2, 4, 6, and 8 under a 10-bit ADC model. To our knowledge, this is the first study to evaluate an LSTM predictor within a predictive-quantization ADC loop that explicitly models saturation.

\item \textbf{SR as the primary optimization target for PQ predictors.} We use SR, the fraction of samples exceeding the residual ADC range, to quantify the frequency of irreversible clipping. OER complements SR by weighting each event by its squared overflow depth.

\item \textbf{A resource-constrained algorithm selection guide.} We compare predictor complexity with reconstruction performance and identify configurations suited to logic-only and microcontroller-assisted implementations.
\end{enumerate}

% ======================================================================
\section{System Model and Predictor Design}
% ======================================================================

\subsection{Predictive Quantization System Model}

We first define the target predictive ADC as a discrete-time closed-loop system operating at sampling period $T_s = 1/f_s$. The subsequent benchmark uses an open-loop form of this model to isolate intrinsic predictor performance. Let $x[n]$ denote the analog input voltage at sample index $n$, quantized by a nominal $N_{\text{total}}$-bit SAR ADC with full-scale range $[V_{\min}, V_{\max}]$ and least significant bit (LSB)
\begin{equation}
\Delta = \frac{V_{\max} - V_{\min}}{2^{N_{\text{total}}}}.
\end{equation}

In the target closed-loop architecture, a predictor $f(\cdot)$ consumes a history of $L$ past reconstructed samples and produces a one-step-ahead estimate:
\begin{equation}
\hat{x}[n] = f\big(y[n-1], y[n-2], \dots, y[n-L]\big),
\end{equation}
where $y[n-1]$ is the reconstructed sample from the previous cycle. The prediction residual is
\begin{equation}
r[n] = x[n] - \hat{x}[n].
\end{equation}

Rather than quantizing $x[n]$ directly over the full $N_{\text{total}}$-bit range, the ADC quantizes only $r[n]$ using a reduced $B_r$-bit stage, where $B_r < N_{\text{total}}$. The quantized residual is
\begin{equation}
\hat{r}[n] = \Delta \cdot \text{clip}\!\left(\text{round}\!\left(\frac{r[n]}{\Delta}\right),\; -2^{B_r-1},\; 2^{B_r-1} - 1\right),
\end{equation}
and the reconstructed sample is
\begin{equation}
y[n] = \hat{x}[n] + \hat{r}[n].
\end{equation}

In the target architecture, $y[n]$ is returned to the predictor history buffer. In the evaluated open-loop protocol, $x[n]$ replaces $y[n]$ in that buffer. This substitution prevents quantization and clipping errors from propagating into later predictions. We use $N_{\text{total}}/B_r$ as a conversion-depth reduction factor; it does not by itself quantify circuit-level switching energy.

\subsection{Residual Saturation Model}

The residual ADC has a finite input range: any residual magnitude exceeding
\begin{equation}
R_{\max} = (2^{B_r-1} - 1) \cdot \Delta
\end{equation}
is clipped. Formally, the clipped residual is
\begin{equation}
r_{\text{clip}}[n] = \begin{cases}
R_{\max}, & r[n] > R_{\max},\\
-R_{\max}, & r[n] < -R_{\max},\\
r[n], & \text{otherwise}.
\end{cases}
\end{equation}

\begin{figure*}[!t]
\centering
\includegraphics[width=5in]{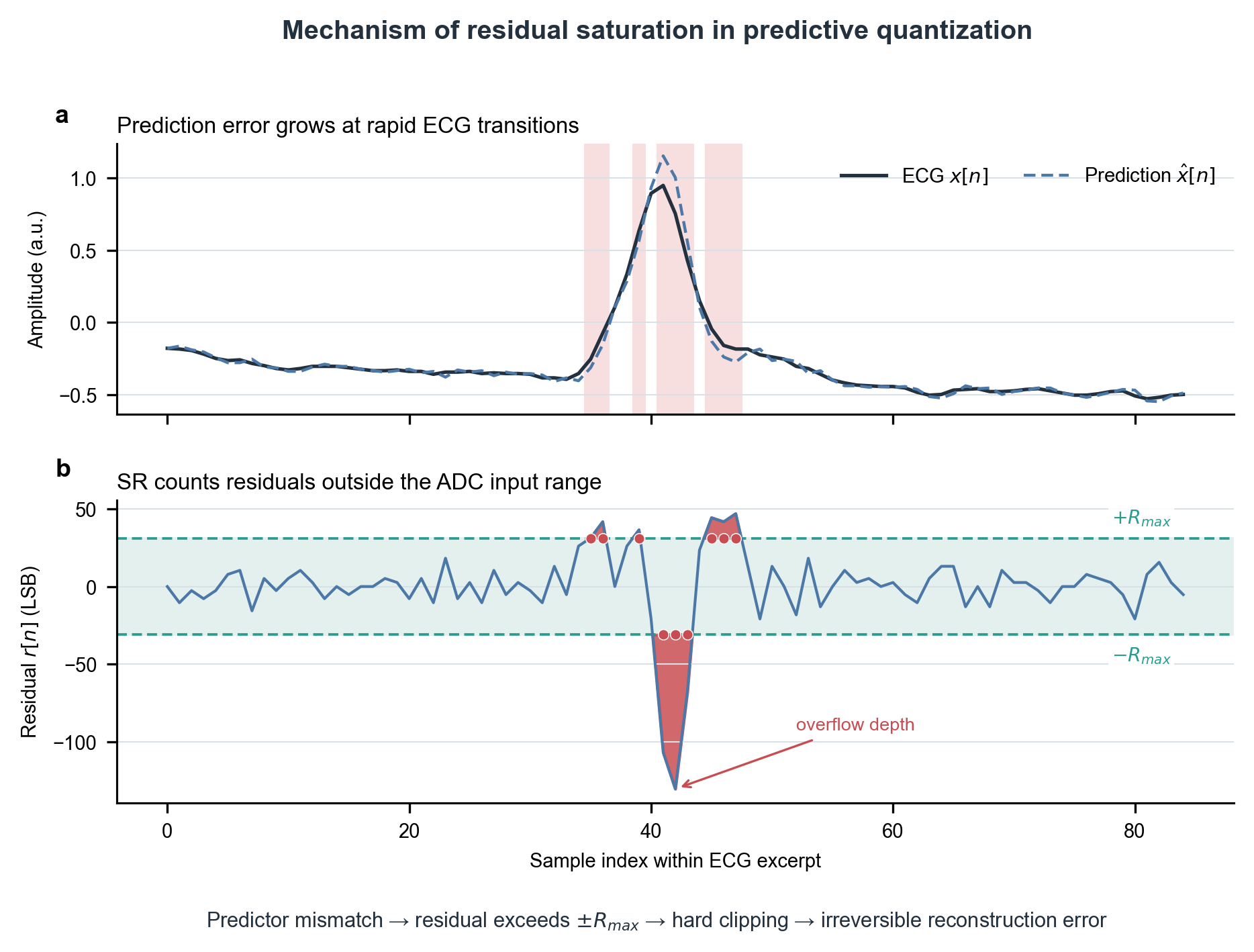}
\caption{Mechanism of residual saturation in predictive quantization at $B_r=6$. (a) A representative ECG excerpt and its one-step-ahead FOT prediction. The predictor tracks slowly varying samples but deviates around the rapid QRS transition; pale-red bands identify samples whose residual exceeds the ADC input range. (b) The corresponding residual in 10-bit LSB units. Dashed lines mark $\pm R_{\max}$, red markers identify saturated samples, and red filled regions show the overflow depth that is removed by hard clipping. SR counts the fraction of samples outside the valid residual range, while overflow-based metrics quantify how far those samples exceed the boundary.}
\label{fig:sr_overflow_mechanism}
\end{figure*}

The clipping operation is nonlinear and irreversible: information in the clipped portion is permanently lost. Figure~\ref{fig:sr_overflow_mechanism} illustrates the causal mechanism using a representative ECG excerpt. Prediction mismatch grows around the rapid QRS transition (Fig.~\ref{fig:sr_overflow_mechanism}a), producing residual samples outside $\pm R_{\max}$ (Fig.~\ref{fig:sr_overflow_mechanism}b). Each threshold crossing contributes one event to $N_{\text{sat}}$, whereas the distance beyond the threshold defines the overflow depth. We define six performance and saturation metrics in Table~\ref{tab:metrics}.

\begin{table*}[!t]
\caption{Residual Saturation Evaluation Metrics}
\label{tab:metrics}
\centering
\begin{tabular}{cclc}
\toprule
\textbf{Metric} & \textbf{Symbol} & \textbf{Definition} & \textbf{Units} \\
\midrule
Signal-to-Noise Ratio & SNR & $10 \log_{10}(\mathbb{E}[x^2] / \mathbb{E}[(x-y)^2])$ & dB \\
Compression Ratio & CR & $N_{\text{total}} / B_r$ & -- \\
Saturation Rate & SR & $(N_{\text{sat}} / N) \times 100$ & \% \\
Mean Overflow Magnitude & MOM & $\frac{1}{N_{\text{sat}}} \sum_{i \in \mathcal{S}} (|r[i]| - R_{\max}) / \Delta$ & LSB \\
Peak Overflow Index & POI & $\max_{i \in \mathcal{S}} (|r[i]| - R_{\max}) / \Delta$ & LSB \\
Overflow Energy Ratio & OER & $\sum_{i \in \mathcal{S}} (|r[i]| - R_{\max})^2 / \sum_n r[n]^2 \times 100$ & \% \\
\bottomrule
\end{tabular}
\end{table*}

where $\mathcal{S}$ is the set of saturated sample indices. OER weights each event by its squared overflow depth. It therefore distinguishes deep clipping near high-energy waveform transitions from shallow threshold crossings.

\subsection{Predictor Design}

We evaluate four predictors spanning the complexity spectrum from zero-parameter linear extrapolation to learned nonlinear sequence models. For each predictor, we describe the problem it addresses relative to simpler alternatives, provide its mathematical definition, quantify its hardware cost, and locate it on the accuracy--complexity tradeoff. The equations below use reconstructed samples $y[n]$ to define the target closed-loop implementation. During the open-loop benchmark, each occurrence of a history sample $y[n-i]$ was replaced by the corresponding original sample $x[n-i]$.

\subsubsection{First-Order Taylor Predictor (FOT)}

In a heavily oversampled regime---ECG at 360\,Hz has a Nyquist frequency of 180\,Hz, yet the clinical bandwidth of interest is below 100\,Hz---adjacent samples are nearly collinear. The FOT exploits this local linearity at the minimum possible hardware cost: it requires no parameters, no training, and no multiplier, establishing the lower bound on both complexity and prediction accuracy.

The predictor extrapolates linearly from the two most recent reconstructed samples:
\begin{equation}
\hat{x}_{\text{FOT}}[n] = 2y[n-1] - y[n-2].
\end{equation}
This follows from truncating the Taylor expansion $x(t) = x(t-\Delta t) + \dot{x}(t-\Delta t)\Delta t + \frac{1}{2}\ddot{x}(t-\Delta t)\Delta t^2 + \cdots$ after the linear term, with the derivative approximated by the backward difference $\dot{x} \approx (y[n-1] - y[n-2])/T_s$ and $\Delta t = T_s$. The truncation error is $O(T_s^2)$ and proportional to the local second derivative. During slowly varying segments (baseline, ST segment, T-wave), the second derivative is small and the approximation is accurate. At QRS onset and the R-wave peak, where curvature is maximal, the truncation error dominates, producing large residuals that are likely to saturate the residual ADC.

The FOT maintains a two-element shift register storing $y[n-1]$ and $y[n-2]$. Each prediction requires one left-shift (multiplication by 2) and one subtraction---no multiplier, no memory beyond the two registers, and no control logic beyond a single clocked update. It defines the lower-left corner of the accuracy--complexity Pareto plot: any predictor claiming practical utility must deliver a meaningful improvement in reconstruction fidelity at an acceptable increase in hardware cost.

\subsubsection{Adaptive-Order Predictor (AOP)}

The FOT commits to a first-order model everywhere, but the ECG signal is non-stationary: it is nearly flat during the isoelectric baseline, approximately linear during the T--P interval and ST segment, and strongly curved through the QRS complex. A predictor that detects the local signal regime and selects the appropriate model order---zeroth-order hold in flat regions, first-order extrapolation in linear-ramp regions, and second-order (quadratic) extrapolation through inflection points---would track the QRS complex more faithfully than the FOT while avoiding the noise amplification that a fixed high-order predictor would incur during flat segments.

At each sample $n \ge 3$, the AOP computes the first and second backward differences of the reconstructed signal:
\begin{equation}
d_1[n] = y[n-1] - y[n-2], \qquad d_2[n] = y[n-1] - 2y[n-2] + y[n-3].
\end{equation}
The first difference $|d_1[n]|$ measures local slope; the second difference $|d_2[n]|$ measures local curvature. The predictor selects the extrapolation order by comparing these quantities against thresholds that scale with the full signal dynamic range $V_{\text{range}} = V_{\max} - V_{\min}$, testing for high curvature first and defaulting to a hold when neither slope nor curvature is significant:
\begin{equation}
\hat{x}_{\text{AOP}}[n] = \begin{cases}
3y[n-1]-3y[n-2]+y[n-3], & |d_2[n]|>\epsilon_1,\\[3pt]
2y[n-1]-y[n-2], & \begin{gathered}|d_2[n]|\le\epsilon_1,\\ |d_1[n]|>\epsilon_0\end{gathered},\\[3pt]
y[n-1], & \text{otherwise}.
\end{cases}
\end{equation}
where $\epsilon_0 = \alpha \cdot V_{\text{range}}$ and $\epsilon_1 = \beta \cdot V_{\text{range}}$. The coefficients $\{1, 2, 3\}$ in the three branches are the Lagrange extrapolation weights for 0th-, 1st-, and 2nd-order polynomial fits through the most recent one, two, and three samples, respectively. The hyperparameters $\alpha \in [5\times10^{-4}, 5\times10^{-2}]$ and $\beta \in [10^{-4}, 5\times10^{-2}]$ are tuned by grid search on the training set, with saturation rate (SR) as the optimization target.

In hardware, the AOP requires only one additional register (three total, for $y[n-1]$, $y[n-2]$, $y[n-3]$) and two comparators beyond the FOT baseline. The arithmetic path uses at most two shifts and two additions per prediction; the coefficients $\{1,2,3\}$ are realized as shifts and adds without a hardware multiplier. The order-decision logic adds two magnitude comparisons per sample. The AOP thus occupies the ``lightweight but adaptive'' point on the Pareto frontier: its hardware cost is nearly indistinguishable from the FOT, yet it substantially reduces prediction error during QRS complexes, where the FOT incurs its deepest saturations. It is the recommended choice when the hardware budget is limited to pure combinatorial logic and a small number of registers.

\subsubsection{Kalman Filter Predictor (KFP)}

Both the FOT and the AOP are deterministic: they treat every past reconstructed sample as equally trustworthy and compute predictions from a fixed formula. Real ECG signals, however, are corrupted by physiological and environmental noise---muscle artifact, motion-induced baseline wander, electrode contact variation, and 50/60\,Hz mains interference. A predictor that maintains an explicit estimate of the signal state together with its uncertainty can optimally fuse a process model of how the signal evolves with the noisy observation, suppressing noise-driven prediction errors that would otherwise cause spurious saturation events. The Kalman filter provides this capability within a principled Bayesian framework.

We model the ECG sample sequence with a constant-velocity (CV) state-space model whose state vector $\mathbf{s}_k = [s_k, \dot{s}_k]^\top$ comprises the signal amplitude and its first derivative at discrete index $k$:
\begin{equation}
\mathbf{s}_{k+1} = \underbrace{\begin{bmatrix} 1 & 1 \\ 0 & 1 \end{bmatrix}}_{\mathbf{F}} \mathbf{s}_k + \mathbf{w}_k, \qquad
z_k = \underbrace{\begin{bmatrix} 1 & 0 \end{bmatrix}}_{\mathbf{H}} \mathbf{s}_k + v_k,
\end{equation}
where $\mathbf{w}_k \sim \mathcal{N}(0, \mathbf{Q})$ and $v_k \sim \mathcal{N}(0, R)$. The process noise covariance is parameterized by the scalar intensity $Q$ using the standard discretization of a continuous white-noise acceleration model:
\begin{equation}
\mathbf{Q} = \begin{bmatrix} Q/4 & Q/2 \\ Q/2 & Q \end{bmatrix}.
\end{equation}

At each index $k$, the filter first \textit{predicts}: the prior state estimate and its covariance are propagated as $\hat{\mathbf{s}}_{k|k-1} = \mathbf{F}\hat{\mathbf{s}}_{k-1|k-1}$ and $\mathbf{P}_{k|k-1} = \mathbf{F}\mathbf{P}_{k-1|k-1}\mathbf{F}^\top + \mathbf{Q}$. The one-step-ahead prediction supplied to the ADC is the first component of the prior state: $\hat{x}_{\text{KFP}}[k] = \hat{s}_{k|k-1}$. The filter then \textit{updates}: the current true sample $z_k = x[k]$ is incorporated via the Kalman gain $\mathbf{K}_k = \mathbf{P}_{k|k-1}\mathbf{H}^\top(\mathbf{H}\mathbf{P}_{k|k-1}\mathbf{H}^\top + R)^{-1}$, yielding the posterior $\hat{\mathbf{s}}_{k|k} = \hat{\mathbf{s}}_{k|k-1} + \mathbf{K}_k(z_k - \mathbf{H}\hat{\mathbf{s}}_{k|k-1})$. Because the observation is scalar, the matrix inversion reduces to a scalar reciprocal.

The two tunable parameters encode prior beliefs about signal dynamics. A small $Q$ asserts that the signal evolves smoothly (appropriate during baseline and ST segments); a large $Q$ allows the filter to track rapid QRS transients. A small $R$ expresses high confidence in the observation; a large $R$ applies heavier smoothing. Both are selected by grid search over $Q \in [0.01, 10]$ and $R \in [0.01, 1]$ on the training set.

The KFP stores six floating-point (FP) values: the $2 \times 1$ state vector ($s_k$, $\dot{s}_k$) and the symmetric $2 \times 2$ covariance matrix $\mathbf{P}$ (three independent entries, stored as four for convenience). Each predict--update cycle requires approximately 10 multiply--accumulate (MAC) operations, dominated by the $2 \times 2$ matrix multiplications, plus one scalar reciprocal. This is feasible on any low-power embedded processor such as an ARM Cortex-M0 or M4. The KFP's explicit noise model makes it the predictor of choice when the input signal is expected to be noisy---for instance, during ambulatory monitoring with motion artifact---and when a small microcontroller-class processor is already present in the signal chain. Its Bayesian foundation also supports future extensions, including online adaptation of $Q$ and $R$ (Section~IV).

\subsubsection{LSTM Predictor}

The FOT, AOP, and KFP all rely on explicit structural assumptions: local linearity, polynomial order switching, or a linear-Gaussian state-space model. ECG signals, however, exhibit nonlinear and non-stationary dynamics that resist closed-form modeling. The QRS morphology varies with heart rate, respiration phase, and electrode placement; the relationship between a sample and its $L$ predecessors depends on the cardiac phase in ways not captured by any fixed-order polynomial or linear dynamical model. A predictor that learns these dependencies directly from data---without specifying the functional form in advance---could anticipate waveform morphology with substantially higher accuracy than any model-based alternative. The LSTM architecture is well-suited to this task because its gated memory cells can retain information over hundreds of time steps while learning to forget irrelevant context.

We use a two-layer stacked LSTM. The input at prediction step $n$ is a window of $L = 72$ past min--max-normalized samples (200\,ms at 360\,Hz, chosen to span slightly more than one QRS duration of $\sim$80--120\,ms):
\begin{equation}
\begin{aligned}
\mathbf{u}_n &=[\tilde{x}[n-L],\ldots,\tilde{x}[n-1]]
                \in \mathbb{R}^{72},\\
\tilde{x}[n] &=\frac{x[n]-x_{\min}}{x_{\max}-x_{\min}}.
\end{aligned}
\end{equation}
The network maps this window through two LSTM layers to a scalar prediction:
\begin{equation}
\mathbf{h}_n^{(1)} = \text{LSTM}^{(1)}(\mathbf{u}_n; \mathbf{W}^{(1)}), \quad \mathbf{h}^{(1)} \in \mathbb{R}^{64},
\end{equation}
\begin{equation}
\mathbf{h}_n^{(2)} = \text{LSTM}^{(2)}(\mathbf{h}_n^{(1)}; \mathbf{W}^{(2)}), \quad \mathbf{h}^{(2)} \in \mathbb{R}^{32},
\end{equation}
\begin{equation}
\hat{\tilde{x}}[n] = \mathbf{w}_{\text{out}}^\top \mathbf{h}_n^{(2)} + b_{\text{out}},
\end{equation}
where the final scalar $\hat{\tilde{x}}[n]$ is inverse-normalized to obtain $\hat{x}_{\text{LSTM}}[n]$. Each LSTM cell implements the standard gate equations:
\begin{equation}
\begin{aligned}
\mathbf{f}_t &= \sigma(\mathbf{W}_f \cdot [\mathbf{h}_{t-1}, \mathbf{u}_t] + \mathbf{b}_f), \quad
\mathbf{i}_t = \sigma(\mathbf{W}_i \cdot [\mathbf{h}_{t-1}, \mathbf{u}_t] + \mathbf{b}_i), \\
\tilde{\mathbf{c}}_t &= \tanh(\mathbf{W}_c \cdot [\mathbf{h}_{t-1}, \mathbf{u}_t] + \mathbf{b}_c), \quad
\mathbf{c}_t = \mathbf{f}_t \odot \mathbf{c}_{t-1} + \mathbf{i}_t \odot \tilde{\mathbf{c}}_t, \\
\mathbf{o}_t &= \sigma(\mathbf{W}_o \cdot [\mathbf{h}_{t-1}, \mathbf{u}_t] + \mathbf{b}_o), \quad
\mathbf{h}_t = \mathbf{o}_t \odot \tanh(\mathbf{c}_t),
\end{aligned}
\end{equation}
where $\sigma(\cdot)$ is the logistic sigmoid and $\odot$ denotes element-wise multiplication. The forget gate $\mathbf{f}_t$ enables the cell to selectively discard stale context when the signal enters a new cardiac phase. Dropout with $p = 0.2$ is applied after each LSTM layer during training only.

The network is trained on the first 80\% of Record 101 to minimize mean squared error (MSE) between $\hat{\tilde{x}}[n]$ and $\tilde{x}[n]$, using Adam ($\eta = 0.001$, $\beta_1 = 0.9$, $\beta_2 = 0.999$) with batch size 32. Training runs for a maximum of 30 epochs with early stopping (patience of 10 epochs on validation loss, using 20\% of the training set as validation). The model achieving the lowest validation loss is retained for evaluation on the held-out test set.

Inference requires approximately 20,000 MAC operations and $\sim$50,000 floating-point parameters ($64 \times 72$ input weights, $4 \times 64 \times 64$ and $4 \times 32 \times 32$ recurrent weights, plus biases and output weights)---three to four orders of magnitude beyond the FOT and AOP. Practical deployment on a wearable device would require model compression: post-training 8-bit quantization (INT8) reduces weight storage by $4\times$ and enables integer-MAC execution on a Cortex-M4 with CMSIS-NN; further compression via weight pruning and knowledge distillation into a smaller student network could bring the cost within the $\sim$100\,$\mu$W range of a low-power MCU. The LSTM occupies the accuracy-ceiling point on the Pareto frontier: it establishes the upper bound on what PQ can deliver with a given amount of temporal context, and it provides a quantitative target for compressed models that seek to approach its accuracy at a fraction of the cost.

\subsection{Dataset and Simulation Procedure}

We evaluated all predictors on a contiguous excerpt from Record 101 of the MIT-BIH Arrhythmia Database~\cite{ref8,ref9}. Record 101 is a two-lead ECG recording sampled at 360\,Hz with 11-bit resolution. We used 5,317 samples ($\sim$14.8\,s) from the modified limb lead II channel. The first 4,253 samples (80\%) were used for training and hyperparameter selection, and the remaining 1,064 samples formed the held-out test set.

The nominal ADC resolution was fixed at $N_{\text{total}}=10$ bits, and each predictor was evaluated at $B_r\in\{2,4,6,8\}$. The benchmark used an open-loop protocol to isolate intrinsic one-step prediction accuracy from recursive reconstruction-error propagation. Each predictor therefore received past original samples rather than the reconstructed samples used by the target closed-loop architecture. This controlled substitution permits a matched comparison of predictor classes but does not reproduce deployment-time error accumulation or encoder--decoder synchronization after saturation. The residual was clipped to $\pm R_{\max}$, quantized at $B_r$ bits, and evaluated using the six metrics in Table~\ref{tab:metrics}. The simulation was implemented in Python using NumPy for FOT, AOP, and KFP, and TensorFlow/Keras for LSTM.

% ======================================================================
\section{Results and Analysis}
% ======================================================================

\subsection{Reconstruction Performance}

Table~\ref{tab:results_4bit} reports reconstruction performance at the reference configuration $B_r=6$, which predicts four MSBs. This setting reduced the nominal conversion depth while keeping SR below 3.3\% for all predictors.

\begin{table}[!t]
\caption{Reconstruction Performance at $N_{\text{total}} = 10$, $B_r = 6$}
\label{tab:results_4bit}
\centering
\begin{tabular}{lccccc}
\toprule
\textbf{Predictor} & \textbf{SNR (dB)} & \textbf{SR (\%)} & \textbf{MOM (LSB)} & \textbf{POI (LSB)} & \textbf{OER (\%)} \\
\midrule
FOT  & 28.39 & 2.69 & 31.66 & 109.70 & 25.29 \\
AOP  & 29.61 & 2.44 & 28.24 & 93.12 & 18.89 \\
KFP  & 30.88 & 2.16 & 21.85 & 79.04 & 12.93 \\
LSTM & 29.28 & 3.27 & 24.82 & 95.44 & 18.38 \\
\bottomrule
\end{tabular}
\end{table}

\begin{figure*}[!t]
\centering
\includegraphics[width=7.0in]{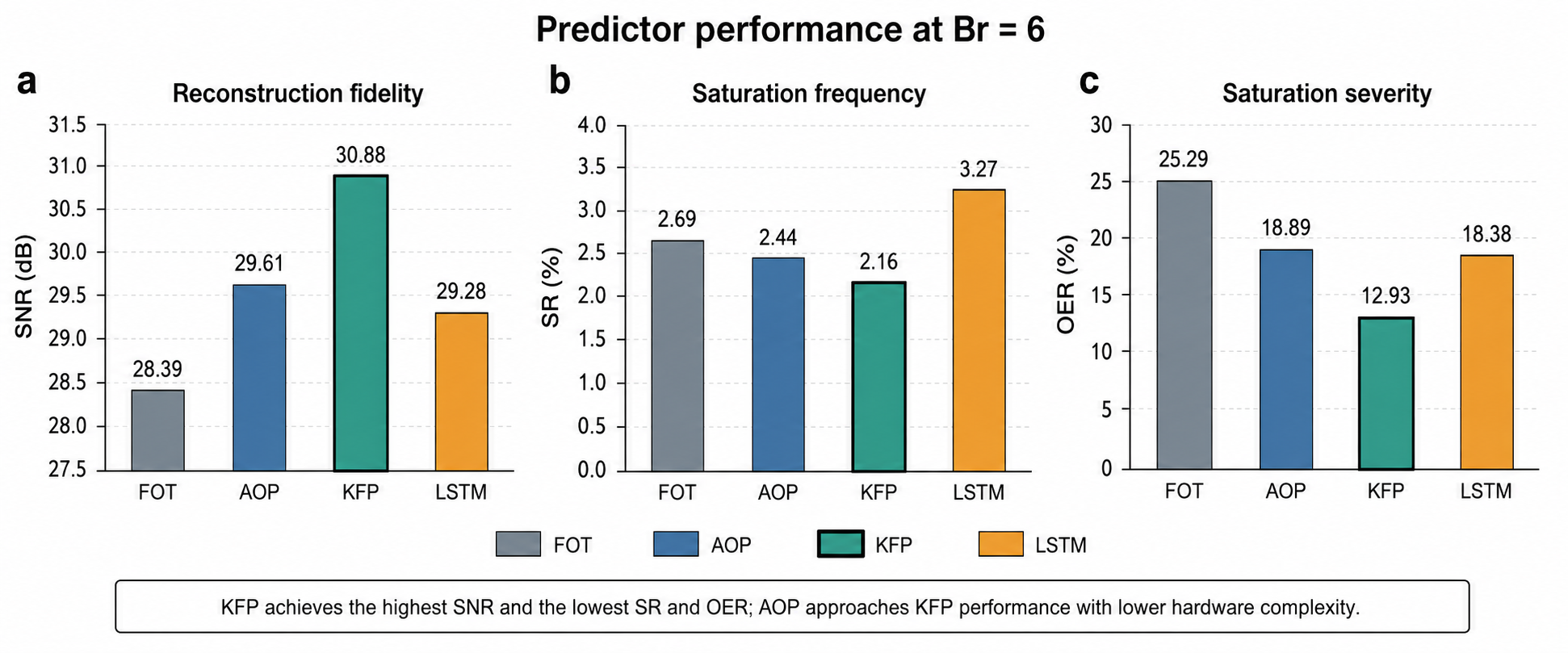}
\caption{Predictor performance at $B_r=6$. (a) Reconstruction fidelity measured by SNR. (b) Saturation frequency measured by SR. (c) Saturation severity measured by OER. KFP achieves the highest SNR and the lowest SR and OER, while AOP approaches its performance with substantially lower hardware complexity. Values are those reported in Table~\ref{tab:results_4bit}.}
\label{fig:predictor_benchmark}
\end{figure*}

Figure~\ref{fig:predictor_benchmark} compares the three principal performance dimensions. KFP achieved the highest SNR (30.88\,dB), lowest SR (2.16\%), and lowest OER (12.93\%). AOP followed with 29.61\,dB SNR, 2.44\% SR, and 18.89\% OER. FOT produced 28.39\,dB SNR and 25.29\% OER, while LSTM produced 29.28\,dB SNR and 3.27\% SR. Thus, the model-based predictors outperformed LSTM on this limited excerpt under the tested configuration.

\subsection{Effect of Residual Bit-Width}

Table~\ref{tab:sr_sweep} reports SR across the full range $B_r \in \{2, 4, 6, 8\}$.

\begin{table}[!t]
\caption{SR (\%) vs.\ Residual Bit-Width}
\label{tab:sr_sweep}
\centering
\begin{tabular}{lcccc}
\toprule
$\boldsymbol{B_r}$ & \textbf{FOT} & \textbf{AOP} & \textbf{KFP} & \textbf{LSTM} \\
\midrule
2 & 86.97 & 85.15 & 92.95 & 91.35 \\
4 & 43.88 & 42.86 & 54.23 & 49.63 \\
6 &  2.69 &  2.44 &  2.16 &  3.27 \\
8 &  0.15 &  0.00 &  0.00 &  0.00 \\
\bottomrule
\end{tabular}
\end{table}

\begin{figure}[!t]
\centering
\includegraphics[width=3in]{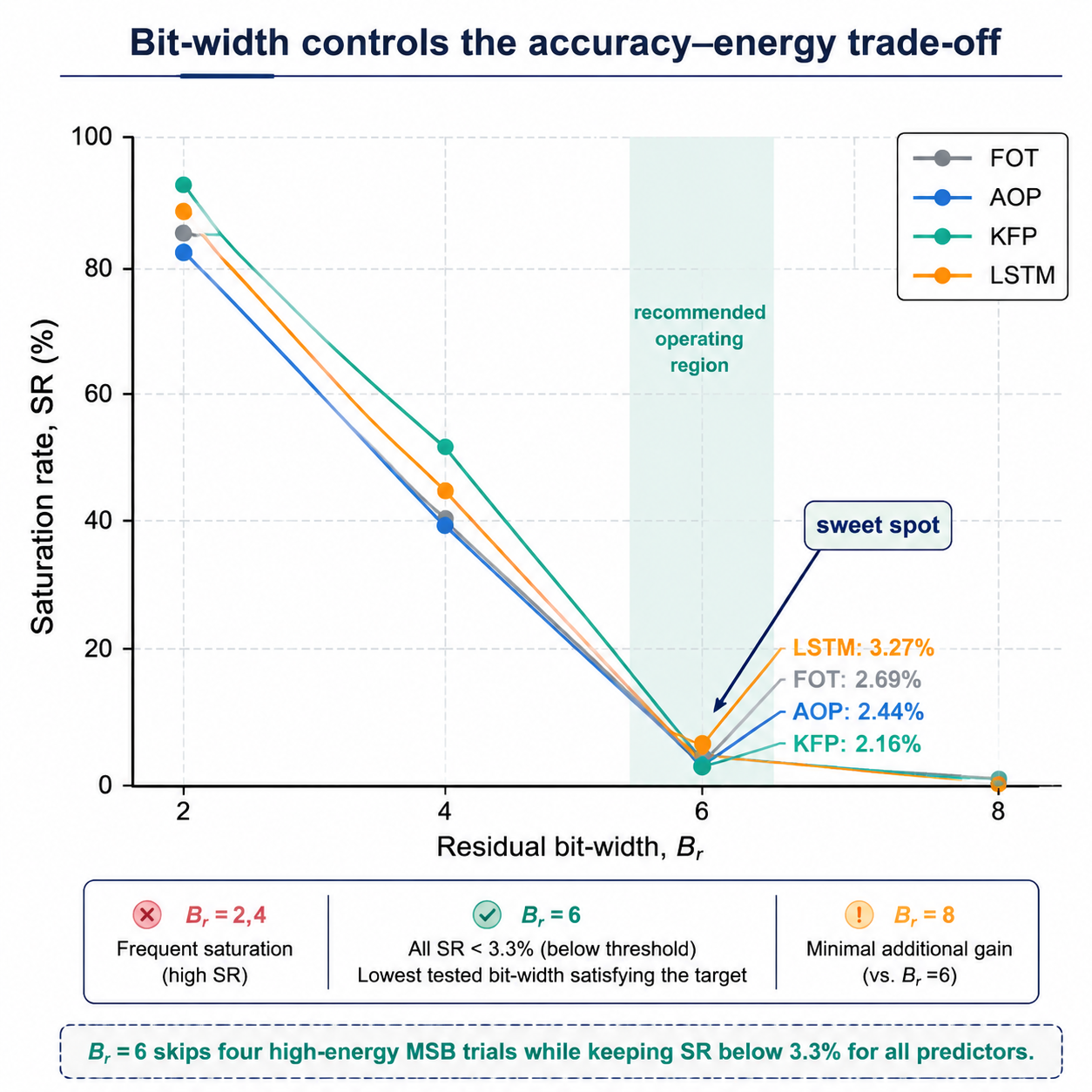}
\caption{Residual bit-width controls the saturation--conversion-depth tradeoff in the open-loop benchmark. Small residual widths ($B_r=2$ and 4) predict more MSBs but produce frequent saturation. At $B_r=6$, SR falls below 3.3\% for all predictors while four MSB trials are omitted in the idealized model. Increasing to $B_r=8$ provides little additional saturation benefit but reduces the conversion-depth reduction.}
\label{fig:accuracy_energy}
\end{figure}

Figure~\ref{fig:accuracy_energy} summarizes the saturation--conversion-depth tradeoff. At $B_r=2$, the symmetric range defined by (8) is $\pm\Delta$. All predictors saturated frequently, with SR above 85\%. KFP had the highest SR at 92.95\%, compared with 86.97\% for FOT. Nevertheless, KFP achieved 22.44\,dB SNR, whereas FOT and AOP achieved 5.21 and 6.10\,dB, respectively.

At $B_r = 4$ (predicting the upper 6 MSBs), SR drops to 42--55\%, and the KFP's SNR advantage persists (24.96\,dB vs.\ 6.10--12.34\,dB). The AOP overtakes the FOT on both SR (42.86\% vs.\ 43.88\%) and SNR (12.34\,dB vs.\ 6.10\,dB), demonstrating the benefit of adaptive-order switching when the residual range is moderately constrained.

At $B_r = 6$ (predicting the upper 4 MSBs), the operating point shifts qualitatively. SR falls below 3.3\% for all predictors, and the SNR spread narrows to a 2.5\,dB range (28.39--30.88\,dB). For the evaluated excerpt, this was the lowest tested width at which saturation became infrequent for the AOP and KFP. The model also omitted four MSB trials, although their circuit-level energy contribution was not measured.

At $B_r = 8$ (predicting the upper 2 MSBs), SR is virtually zero for the AOP, KFP, and LSTM, and only 0.15\% for the FOT. SNR climbs above 48\,dB for all predictors, but only two MSB cycles are omitted. Within the tested open-loop conditions, $B_r = 6$ provided a more conservative balance between fidelity and conversion-depth reduction than $B_r = 4$.

\subsection{Computational Complexity and Hardware Cost}

Predictor complexity provides the second axis of the accuracy--cost tradeoff. FOT requires two registers, one shift, and one subtraction per sample. AOP adds one register and two comparisons, with coefficients implemented through shifts and additions rather than multiplication. At $B_r=6$, this additional logic increased SNR by 1.22\,dB and reduced SR by 0.25 percentage points relative to FOT. Circuit-level power remains to be measured.

KFP requires a state vector, covariance matrix, approximately 10 MACs, and one scalar reciprocal per sample. At 360\,Hz, this corresponds to approximately 3,600 MACs per second. This workload is compatible with microcontroller-class execution, although its incremental energy cost requires hardware measurement.

The LSTM occupies a fundamentally different cost regime: $\sim$20,000 MACs per sample at 360\,Hz translates to 7.2 million operations per second, three orders of magnitude beyond the KFP. Practical deployment would require model compression~\cite{ref12}: post-training 8-bit quantization and weight pruning could substantially narrow this gap, but the compression step is non-trivial and may erode the accuracy advantage that motivates the LSTM in the first place.

% ======================================================================
\section{Discussion}
% ======================================================================

\subsection{Algorithm Selection Guide}

\textbf{Ultra-low-power, logic-only:} AOP is the recommended choice---near-baseline hardware cost, SNR within 1.3\,dB of the best predictor, and SR of 2.44\% at $B_r = 6$. Its hardware overhead over the FOT is one additional register and two comparators.

\textbf{Best overall accuracy:} KFP achieves the highest SNR (30.88\,dB) and lowest SR (2.16\%) at $B_r = 6$, using $\sim$10 MAC/sample---well within the capability of low-power microcontroller units (MCUs). At low $B_r$ values, however, the KFP underperforms even the FOT on SR (Table~\ref{tab:sr_sweep}), revealing a sensitivity to the residual ADC range that designers must account for.

\textbf{Data-rich scenarios with relaxed compute constraints:} LSTM did not outperform the model-based predictors on the available excerpt. Whether broader multi-subject training reverses this ranking remains an open empirical question. Embedded deployment would also require model compression~\cite{ref12}.

\textbf{Operating-point recommendation under the tested conditions:} AOP + $B_r = 6$ (predicting the upper 4 MSBs) provided a favorable accuracy-to-complexity ratio. At this width, the AOP achieved 29.61\,dB SNR and 2.44\% SR while the idealized conversion model omitted four MSB cycles. Circuit-level measurements are required to translate this conversion-depth reduction into energy savings.

\subsection{Predictor--ADC Co-Design Implications}

\textbf{SR as the primary design metric.} SR directly measures the fraction of samples lost to clipping and varied by more than 90 percentage points across the tested widths. SNR combines bounded quantization noise with saturation distortion and therefore cannot identify the source of reconstruction error. We consequently use SR as the primary operational metric, with SNR and OER providing complementary fidelity and severity information. Clinical consequences were not evaluated in this study.

\textbf{Diminishing returns at both extremes.} At $B_r=2$, the range defined by the model is $\pm\Delta$, and every predictor saturated frequently. At $B_r=8$, saturation was nearly absent, but only two MSBs were predicted. The tested data therefore support $B_r=6$ as the lowest width that maintained SR below 3.3\% for all predictors.

\textbf{Overflow-sensitive optimization.} The SR and OER results show that average prediction error alone does not characterize PQ performance. SR measures how often the residual exceeds $R_{\max}$, whereas OER measures the energy removed by clipping. A predictor can therefore achieve competitive SNR while retaining infrequent but severe overflow events. This distinction motivates a saturation-aware training objective that augments the standard MSE loss with a penalty for residuals exceeding the ADC range:
\begin{equation}
\mathcal{L} = \frac{1}{N} \sum_n (x[n] - \hat{x}[n])^2 + \lambda \sum_n \max\!\big(0, |r[n]| - R_{\max}\big)^2,
\end{equation}
where $\lambda$ controls the tradeoff between average fidelity and tail suppression. This loss could be applied when training the LSTM on a larger multi-record corpus, potentially closing the SR gap with the KFP observed in Table~\ref{tab:results_4bit}.

\subsection{Limitations and Future Work}

\textbf{Single-record evaluation.} Results are on MIT-BIH Record 101 only. Multi-record evaluation across the full database (48 records) and independent datasets is needed to establish generalization.

\textbf{Closed-loop operation with overflow detection and mode recovery.} The present benchmark uses past original samples and therefore excludes recursive reconstruction-error propagation. Future work will implement a closed-loop architecture containing an overflow detector and a two-state conversion controller. During predictive-quantization mode, the detector will test whether $|r[n]|>R_{\max}$ before residual clipping. A detected overflow will terminate the current predictive conversion and switch the ADC to conventional full-resolution mode. The resulting full-resolution sample will replace the saturated reconstruction in both the encoder and decoder histories. The ADC will remain in conventional mode for $L$ consecutive samples, where $L$ denotes the predictor history length. This interval flushes every overflow-contaminated entry from the history buffer. For example, a predictor that requires eight historical samples will perform eight conventional conversions before predictive quantization resumes. A shared mode indicator and synchronized buffer update will be required to prevent encoder--decoder divergence. Stateful predictors, including KFP and LSTM, may additionally require state reinitialization from the refreshed sample sequence. Future evaluation will quantify detection and recovery latency, fallback frequency, conversion-depth overhead, and residual error propagation. It will also determine whether the predictor rankings reported here remain stable under the proposed recovery mechanism.

\textbf{Simulation-only validation.} Behavioral abstraction assumes ideal DAC linearity, no comparator noise, and no timing errors. RTL-level simulation and test-chip measurement are required for silicon validation.

\textbf{No adaptive Q/R for Kalman.} Fixed noise parameters limit tracking of non-stationary ECG statistics. Revach et al.~\cite{ref10} proposed online expectation-maximization (EM) learning of noise covariances; Avenda\~{n}o et al.~\cite{ref11} used forgetting-factor-based recursive estimation. Integrating these into the PQ loop is a natural extension.

\textbf{Full-precision LSTM.} Deployment requires quantization. Dynamically-biased LSTM (DB-LSTM)~\cite{ref12} has demonstrated $>98\%$ accuracy retention with 4-bit (INT4) and 3-bit (INT3) quantization on edge-AI hardware.

\textbf{Generalization beyond ECG.} The PQ framework applies to any band-limited, temporally correlated signal (EEG, EMG, PPG, audio).

\textbf{KFP anomaly at low residual bit-widths.} KFP had higher SR than FOT at $B_r=2$ (92.95\% versus 86.97\%) and $B_r=4$ (54.23\% versus 43.88\%). Because the benchmark was open loop, this behavior cannot be attributed to clipped samples propagating through the predictor state. It may instead reflect the shape of the KFP residual distribution, initialization transients, or parameters selected for a different operating regime. Distinguishing these explanations requires transient removal, width-specific tuning, and residual-tail analysis.

% ======================================================================
\section{Conclusion}
% ======================================================================

We compared four one-step-ahead predictors for a 10-bit PQ ECG model using residual widths from 2 to 8 bits. The benchmark used a 5,317-sample excerpt from MIT-BIH Record 101 and an open-loop, saturation-aware simulation.

At $B_r=6$, KFP achieved the highest reconstruction quality, with 30.88\,dB SNR, 2.16\% SR, and 12.93\% OER. AOP achieved 29.61\,dB SNR and 2.44\% SR using three registers, shift-and-add arithmetic, and two comparators. OER further distinguished deep overflow events from shallow threshold crossings. LSTM did not outperform the model-based predictors on this limited excerpt.

Under the evaluated excerpt and open-loop protocol, AOP with $B_r=6$ provided a favorable accuracy--complexity configuration, while KFP provided the highest fidelity. These findings do not establish closed-loop stability, circuit-level energy savings, or population-level generalization. Future work will evaluate reconstructed-sample feedback and an overflow-triggered fallback to full-resolution conversion. It will also examine recovery after history-buffer refresh, multi-subject performance, and hardware energy consumption before deployment claims are made.

% ======================================================================
\section*{Acknowledgments}
% ======================================================================

The authors would like to thank the MIT-BIH Arrhythmia Database team for maintaining and distributing the benchmark dataset used in this study.

% ======================================================================

% ======================================================================
% Biography (placeholder)
% ======================================================================

\vspace{11pt}

\begin{IEEEbiographynophoto}{First A. Author}
Biography text here.
\end{IEEEbiographynophoto}

\begin{IEEEbiographynophoto}{Second B. Author}
Biography text here.
\end{IEEEbiographynophoto}

\begin{IEEEbiographynophoto}{Third C. Author}
Biography text here.
\end{IEEEbiographynophoto}

\vfill


\begin{thebibliography}{28}
\bibliographystyle{IEEEtran}

\bibitem{ref1}
C.~Feng, C.~Liu, Y.~Wu, and H.~Qian, ``A predictive quantization based compressive sensing SAR ADC for ECG signal,'' \textit{Analog Integr. Circ. Signal Process.}, vol.~124, p.~25, May 2025.

\bibitem{ref2}
X.~Tang, M.~Renteria-Pinon, and W.~Tang, ``Dynamic predictive sampling analog to digital converter for sparse signal sensing,'' \textit{IEEE Trans. Circuits Syst. II}, vol.~70, no.~6, pp.~2360--2364, Jun. 2023.

\bibitem{ref3}
M.~Renteria-Pinon, X.~Tang, and W.~Tang, ``Fully digital second-order level-crossing sampling ADC for data saving in sensing sparse signals,'' \textit{IEEE Trans. Biomed. Circuits Syst.}, vol.~18, 2024.

\bibitem{ref3b}
Z.~Zhang, J.~Li, Q.-H.~Zhang, N.~Ning, and Q.~Yu, ``A 10-bit fully-predictive ADC with code-recombination algorithm in transducing sensor node signals,'' in \textit{Proc. IEEE Int. Symp. Circuits Syst. (ISCAS)}, 2020.

\bibitem{ref4}
K.~I.~Arafa, D.~M.~Ellaithy, A.~Zekry, M.~Abouelatta, and H.~Shawkey, ``Successive approximation register analog-to-digital converter (SAR ADC) for biomedical applications,'' \textit{Active Passive Electron. Compon.}, vol.~2023, pp.~1--29, Jan. 2023.

\bibitem{ref5}
X.~Tong and M.~Ghovanloo, ``Energy-efficient switching scheme in SAR ADC for biomedical electronics,'' \textit{Electron. Lett.}, vol.~51, no.~4, pp.~315--317, Feb. 2015.

\bibitem{ref6b}
J.-Y.~Lin, C.-C.~Hsieh, W.-H.~Chang, H.-H.~Tsai, and C.-F.~Chiu, ``A 9.2b 47fJ/conversion-step asynchronous SAR ADC with input range prediction DAC switching,'' in \textit{Proc. IEEE Int. Symp. VLSI Design, Automation and Test (VLSI-DAT)}, 2018.

\bibitem{ref6}
H.~Zacarias, R.~Martins, J.~Rodrigues, and J.~M.~R.~S.~Tavares, ``ECG forecasting system based on long short-term memory,'' \textit{Bioengineering}, vol.~11, no.~1, art.~77, Jan. 2024.

\bibitem{ref7}
S.~Wang, J.~Li, H.~Zhang, and Y.~Liu, ``Early prediction of sudden cardiac death risk with nested LSTM based on electrocardiogram sequential features,'' \textit{BMC Med. Inform. Decis. Mak.}, vol.~24, art.~102, Apr. 2024.

\bibitem{ref8b}
Y.~Zhang, Z.~Wang, M.~Zhao, and Z.~Tan, ``A 20kHz 87.2dB SNDR 12.1$\mu$W NS SAR ADC based on quantization-error-matched RLSB-first algorithm,'' in \textit{Proc. IEEE Asian Solid-State Circuits Conf. (A-SSCC)}, 2023.

\bibitem{ref8}
G.~B.~Moody and R.~G.~Mark, ``The impact of the MIT-BIH Arrhythmia Database,'' \textit{IEEE Eng. Med. Biol. Mag.}, vol.~20, no.~3, pp.~45--50, May/Jun. 2001.

\bibitem{ref9}
A.~L.~Goldberger \textit{et al.}, ``PhysioBank, PhysioToolkit, and PhysioNet: Components of a new research resource for complex physiologic signals,'' \textit{Circulation}, vol.~101, no.~23, pp.~e215--e220, Jun. 2000.

\bibitem{ref10}
G.~Revach \textit{et al.}, ``HKF: Hierarchical Kalman filtering with online learned evolution priors for adaptive ECG denoising,'' \textit{IEEE Trans. Signal Process.}, vol.~72, pp.~4084--4098, 2024.

\bibitem{ref11}
L.~E.~Avenda\~{n}o, C.~G.~Castellanos, and J.~I.~Mart\'{i}nez, ``Dual Kalman filter for ECG denoising,'' \textit{Ing. Investig.}, vol.~39, no.~1, pp.~53--61, 2019.

\bibitem{ref12}
S.~Kim and J.~Park, ``DB-LSTM: Dynamically-biased LSTM for time-series analysis on Edge-AI hardware for healthcare monitoring,'' \textit{arXiv preprint}, arXiv:2504.15178, Apr. 2025.

\bibitem{ref13}
J.~Van~Rethy, M.~De~Smedt, M.~Verhelst, and G.~Gielen, ``Predictive sensing in analog-to-digital converters for biomedical applications,'' in \textit{Proc. IEEE Int. Symp. Circuits Syst. (ISCAS)}, 2013.

\bibitem{ref14}
N.~Wood and N.~Sun, ``Predicting ADC: A new approach for low power ADC design,'' in \textit{Proc. IEEE Dallas Circuits Syst. Conf. (DCAS)}, 2014.

\bibitem{ref15}
M.~Namavar, R.~Lotfi, and A.~M.~Sodagar, ``A 10-b 330nW third-order predictive SAR ADC dedicated to neural recording brain implants,'' in \textit{Proc. IEEE Biomed. Circuits Syst. Conf. (BioCAS)}, 2020.

\bibitem{ref16}
F.~M.~Yaul and A.~P.~Chandrakasan, ``A 10 bit SAR ADC with data-dependent energy reduction using LSB-first successive approximation,'' \textit{IEEE J. Solid-State Circuits}, vol.~51, no.~12, pp.~2828--2838, Dec. 2016.

\bibitem{ref17}
J.~Chen, X.~Liu, C.~Yang, J.~Jin, Z.~Chang, and J.~Zhou, ``Predictive LSB-first successive approximation for SAR analog-to-digital converters,'' in \textit{Proc. IEEE Int. Conf. Electron Devices Solid-State Circuits (EDSSC)}, 2022.

\bibitem{ref18}
J.-Y.~Lin and C.-C.~Hsieh, ``A 0.3\,V 10-bit SAR ADC with first 2-bit guess in 90-nm CMOS,'' \textit{IEEE Trans. Circuits Syst. I}, vol.~64, no.~3, pp.~556--568, Mar. 2017.

\bibitem{ref19}
B.~Canal, H.~D.~Klimach, S.~Bampi, and T.~R.~Balen, ``Time assisted SAR ADC with bit-guess and digital error correction,'' in \textit{Proc. IEEE Int. Symp. Circuits Syst. (ISCAS)}, 2022.

\bibitem{ref20}
A.~Scanlan, D.~O'Hare, M.~Halton, V.~O'Brien, B.~Mullane, and E.~Thompson, ``Analysis of feedback predictive encoder based ADCs,'' \textit{Analog Integr. Circ. Signal Process.}, vol.~100, pp.~537--550, 2019.

\bibitem{ref20b}
H.-W.~Lu, X.-P.~Yu, Z.-H.~Lu, K.-S.~Yeo, and J.-M.~Chen, ``A data-dependent energy reduction algorithm for SAR ADC using self-adaptive window,'' \textit{IEEE Trans. Circuits Syst. II}, vol.~68, no.~6, pp.~1832--1836, Jun. 2021.

\bibitem{ref21}
Z.~Zhang, J.~Li, Q.~Zhang, K.~Wu, N.~Ning, and Q.~Yu, ``A dynamic tracking algorithm based SAR ADC in bio-related applications,'' \textit{IEEE Trans. Circuits Syst. II}, vol.~68, no.~7, pp.~2468--2472, Jul. 2021.

\bibitem{ref22}
J.~R.~Shakya, A.~K.~Vadakkan, and G.~C.~Temes, ``Predictive noise shaping SAR ADC,'' in \textit{Proc. IEEE Int. Symp. Circuits Syst. (ISCAS)}, 2022.

\bibitem{ref23}
J.~Mitrovic, Y.~Zhang, and Z.~Ignjatovic, ``Predictive successive approximation ADC,'' in \textit{Proc. IEEE Int. Midwest Symp. Circuits Syst. (MWSCAS)}, 2020.

\bibitem{ref24}
R.~Inanlou, M.~Safarpour, and O.~Silven, ``Arithmetic tracking adaptive SAR ADC for signals with low-activity periods,'' \textit{IEEE Trans. Circuits Syst. I}, vol.~69, no.~9, pp.~3556--3567, Sep. 2022.

\end{thebibliography}
\end{document}